\documentclass[12pt]{article}
\usepackage{graphicx,epsf,amsmath,amsfonts,amssymb,amsbsy}

\input epsf

\font\eulerm=eufm10 scaled \magstep1
\def\a{\mbox{\eulerm a}}
\def\b{\mbox{\eulerm b}}
\def\c{\mbox{\eulerm c}}
\def\u{\mbox{\eulerm u}}
\font\eulermi=eufm7 scaled \magstep1
\def\ai{\mbox{\eulermi a}}
\def\bi{\mbox{\eulermi b}}
\def\ci{\mbox{\eulermi c}}
\def\ui{\mbox{\eulermi u}}

\def\i{\mbox{i}}
\def\\i{\mbox{\scriptsize{i}}}
\def\d{\mbox{d}}
\def\Im{\mbox{Im}}
\def\Re{\mbox{Re}}

\title{Contributions of unstable and virtual particles to nuclear form
factors }
\author{M. L. Nekrasov\\
{\small\it 
Institute for High Energy Physics, NRC ``Kurchatov
Institute'',}  \vspace*{-4\baselineskip}\\
{\small\it Protvino 142281, Russia} }
\date{}
\begin{document}
\maketitle
\begin{abstract}
The coherent meson scattering off heavy nuclei with the production of
two particles in the final state is investigated. We obtain the form
factors for the direct production of the final state and through
intermediate particles, unstable ones that can decay inside the nucleus
and virtual ones. The cases of scattering both in the Coulomb and in the
strong field of the nucleus are considered. This work is stimulated by
an experimental study of the chiral anomaly in a beam of charged kaons
in the OKA facility.
\end{abstract}

\section{Introduction}\label{sec1}

Reactions of coherent production off nuclei provide important
information on the behavior of hadron systems in a nuclear environment
and simultaneously on hadron interactions that are difficult or
impossible to study in other ways. In particular, they provide a unique
opportunity to measure vertices with anomalous parity of the type $K K
\pi \gamma$ or $\pi\pi\pi \gamma$, predicted by the effective
Wess-Zumino-Witten action \cite{Wess-Zumino,Witten}. These vertices are
not available for study in the decay reactions, and their measurement is
obstructed by large backgrounds in the proton-target experiments.
However, they become available in coherent scattering of charged kaons
and pions off nuclei with large atomic numbers owing to the factor
$Z^2$, where $Z$ is the nuclear charge. 

Specifically, we mean reactions of coherent production of $K^{\pm}
\pi^0$ or $\pi^{\pm}\pi^0$ off heavy nuclei in the $K^{\pm}$ or
$\pi^{\pm}$ beams. At sufficiently high energies and low momentum
transfers the Coulomb contributions in these reactions are dominant and
can be separated from competing strong contributions \cite{BermanDrell}.
The most favorable transfer is twice its minimum value, when Coulomb
contributions reach maximum. In turn, among the Coulomb-type
contributions the significance of the anomalous ones increases with
decreasing invariant masses of $K^{\pm} \pi^0$ and $\pi^{\pm}\pi^0$.
Thus, the favorable domain for determining the anomalous vertices is
located at small transfers and small invariant masses of mesons in the
final state \cite{Terentev}. 

\begin{figure}[t]
\hbox{ \hspace*{65pt}
       \epsfxsize=0.65\textwidth \epsfbox{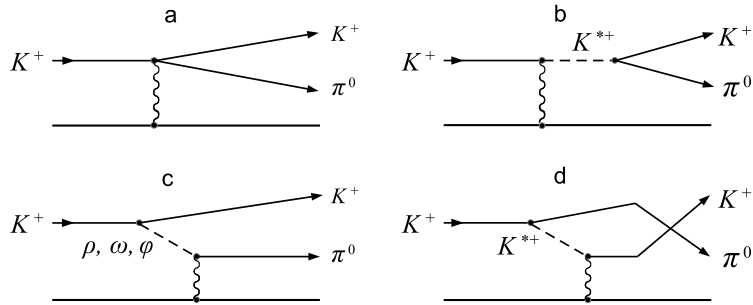}}
\caption{\small Coherent production of $K^{+}\pi^0$ in the $K^{+}$ beam
in the Coulomb field of nucleus: direct production owing to the chiral
anomaly (a), via $K^{*+}$ in the $s$-channel (b), via $\rho,\omega,\phi$
in the $t$-channel (c) and $K^{*+}$ in the $u$-channel (d). The same
diagrams describe similar processes due to strong field of the nucleus
with the replacement of photon by reggeon.
%\vspace*{-0.5\baselineskip} 
}
\label{Fig1}
\end{figure}

In reality, however, the measurements are carried far beyond the
mentioned domain, with the excess in the transfer $t$ by 2 orders of
magnitude \cite{Burtovoy17,Antipov}. As a result, background
contributions become significant. Fig.\ref{Fig1} repre\-sents relevant
diagrams in the case of kaon beam \cite{Rogalyov,Burtovoy13,Vysotsky}.
Preliminary estimates \cite{Vysotsky,Burtovoy13} show that the
background contributions become comparable to the anomaly one shown in
Fig.\ref{Fig1}a. Unfortunately, ambiguities in the form factors prevent
more accurate definition of all the contributions. Point is that the
data for processing were collected in the transfer region that overlaps
the geometric size of the nucleus in configuration space. Specifically,
\cite{Burtovoy17} used data at $|t| \leq 0.025$ (GeV$\!$/c)$^2$ which
means distances 1.2~fm and more with the nucleus radius for Cu target
4.4 fm, and \cite{Antipov} used $|t| \leq 0.025$ (GeV$\!$/c)$^2$ which
means 1.2 fm with the nucleus radius 2.5--4.2 fm for various targets. In
both cases a significant part of inside of the nucleus falls into the
measured region. This should significantly affect the scattering in the
strong field and could affect the Coulomb scattering (because
considerable part of the nuclear surface, where Coulomb forces reach
maximum, falls into the interaction region). In the case of real stable
particles the way to account for these effects is well understood in
Glauber multiple-scattering theory \cite{Glauber1,Glauber2,Faeldt43}.
However, it is not known how to describe the effect in the presence of
intermediate particles such as unstable particles that can decay inside
the nucleus and especially the virtual particles. Accordingly, a method
of determining the form factors in the mentioned cases is  not known,
too.

In this paper we solve this problem. First of all, we define the form
factors in the case of direct production of pairs of particles in the
final state. The solution basically follows the extension of Glauber
theory to the scattering of composite systems \cite{Czyz,Formanek}. The
contributions of intermediate unstable particles we consider on the
basis of their description as superpositions of quasi-stable particles
and their decay products. In the case of virtual particles we use the
field-theoretical analysis of coherent scattering of fast particles off
nonrelativistic ``soft'' systems composed of many constituents
\cite{AAV}. 

We carry out analysis mostly in a general form, but~in the case of
contributions of the anomaly we turn to the particular reaction $K^{+} A
\to K^{+}\pi^{0} A$, currently investigated in the OKA experiment 
(IHEP, Protvino) \cite{OKA}. A theoretical study of this reaction in the
context of the mentioned experiment was carried out in
\cite{Rogalyov,Burtovoy13,Vysotsky}. Unfortunately, any details of the
interaction of the incident particle with the nucleus were not taken
into account, since the form factors were introduced in the unified
Gaussian form. However, the forms factors are much more complex and
their structure depends on the underlying process. In the present paper,
we define the forms factors individually for each of the diagrams
Fig.\ref{Fig1}a,b,c,d. 

In the next section, we describe the method of our study. The form
factor for the direct production of pairs of particles is determined in
sect.~\ref{sec3}. In sect. \ref{sec4} and \ref{sec5} we define the form
factors in the presence of unstable and virtual particles. The results
are discussed in sect.~\ref{sec6}.

\section{Basic approximation}\label{sec2}

The main means for analysis of the collisions of fast particles with
nucleus followed by elastic or quasi-elastic scattering at small angles
is provided by Glauber multiple-scattering theory
\cite{Glauber1,Glauber2}. This theory is based on the assumption that
the nucleons of the nucleus are ``frozen'' during the passage of the
fast particle, and the impact of each of the nucleons on the incident
particle does not depend on the impact of other nucleons. The first
condition means that nucleon of the nucleus can be characterized by
their positions determined by the wave function. The second condition
means additivity of the eikonal phase of the scattered particle.

In this approach, the amplitude of the coherent production of particle
$\a^{*}$ in the $\a$ beam off a nucleus consisting of $A$ nucleons is
determined as follows: \footnote{The given formula defines the leading
approximation with a single inelastic conversion. Hereinafter in this
section we mainly follow \cite{Faeldt43}.} 
\begin{eqnarray}\label{N1}
& \displaystyle
F^{{\ai} \to {\ai}^{*}}\!(\!\vec{\,\bf q}) = 
\int \! \d^3 \vec{\bf r} \; 
e^{\mbox{\scriptsize{i}} {\vec{\bf q} \vec{\bf r}}}
\int \prod_{n=1}^{A} \d \!\vec{\,\bf r}_n 
\left| \Psi(\!\vec{\,\bf r}_1,\dots,\!\vec{\,\bf r}_A) \right|^2
& \nonumber\\
& \displaystyle 
\times \sum_{n=1}^{A} \; \prod_{{j=1}\atop{j\not=n}}^{A-1} \! 
\left[1\!-\! \theta(z_j\!-\!z) 
\gamma^{\ai^{*}}_j \! ({\bf b}\!-\!{\bf b}_j ) \, \right]
f_{n}^{{\ai} \to {\ai}^{*}} \! ( {\vec{\bf r}\!-\!\vec{\bf r}}_n ) 
\prod_{{j=1}\atop{j\not=n}}^{A-1} \!
\left[1\!-\! \theta(z\!-\!z_j) 
\gamma^{\ai}_j ( {\bf b}\!-\!{\bf b}_j ) \, \right] \!. \;&
\end{eqnarray}
Here $\bar{\bf q} = ({\bf q},q_{z})$ is the momentum transferred to the
nucleus with ${\bf q}$ is its two-dimensional component in the
impact-parameter plane and axis $z$ is oriented along the direction of 
motion of the projectile. The longitudinal component is $q_{z} =
(m_{\ai^{*}}^2 - m_{\ai}^2)/(2k)$ where $m_{\ai^{*}}$ and $m_{\ai}$ are
the masses of $\a^{*}$ and $\a$, respectively, $k$ is the $\a$ momentum
in the lab frame. The $\Psi$ is the wave function of the nucleus,
$\bar{\,\bf r}_j = ({\bf b}_j, z_j) $ are the coordinates of the
``frozen'' nucleons counting from the center of mass of the nucleus (we
neglect the effect of the c.m.~motion). The $f_{n}^{{\ai} \to {\ai}^{*}}
\! ( {\bar{\bf r}\!-\!\bar{\bf r}}_n )$ is the amplitude in the
coordinate space of the conversion ${\a} \to {\a}^{*}$ off $n$-th
nucleon, averaged over its isotopic spin and spin (effectively off a
spinless nucleon), $\bar {\,\bf r} = ({\bf b},z)$ is a point where the
conversion $\a \to \a^{*}$ occurs. The profile functions of elastic
scattering before and after the conversion are defined in the standard
way,
\begin{equation}\label{N2}
\gamma_j^{\ai(\ai^{*})} ({\bf b}) = \frac{1}{2\pi \i k} 
\int \d^2 {\bf q} \; 
e^{-\\i{\scriptsize \bf q}{\scriptsize \bf b}} 
f_j^{\,\ai(\ai^{*})}({\bf q})\,,
\end{equation} 
where $f_j^{\,\ai(\ai^{*})}$ is the eikonal amplitude of elastic
scattering of $\a(\a^{*})$ off $j$-th nucleon. All amplitudes are
normalized by $\d\sigma / \d\Omega = |F|^2$ and defined in the lab
frame.

Formula (\ref{N1}) is greatly simplified if we neglect the nucleon
correlations and assume that all nucleons are described by the same
wave functions. In this case
\begin{equation}\label{N3}
|\psi (\vec{\bf r}_1 \ldots \vec{\bf r}_A)|^2 = 
\prod_{n=1}^{A} \rho(\vec{\bf r}_n)\,,
\end{equation}
where $\rho(\vec{\bf r}_n)$ is the distribution density of a single
nucleon,
\begin{equation}\label{N4}
\int \rho(\vec{\bf r}) \; \d ^3 \vec{\bf r} = 1 \,.
\end{equation}
By virtue of (\ref{N3}) and neglecting contributions of order $1/A$, we
can collect the products in (\ref{N1}) to the exponent. Then (\ref{N1})
is reduced to
\begin{equation}\label{N5}
F^{{\ai} \to {\ai}^{*}}\!(\vec{\bf q}) = 
\int \! \d^3 \vec{\bf r} \; 
e^{\mbox{\scriptsize{i}} {\vec{\bf q} \vec{\bf r}}}
\left[ \sum_{n=1}^{A} \; \int \d^3 \vec{\bf r}\,' \,
f_{n}^{{\ai} \to {\ai}^{*}} \! ( {\vec{\bf r}\!-\!\vec{\bf r}\,'} ) 
\rho(\vec{\bf r}\,') \right] 
e^{\mbox{\scriptsize{i}} \chi_{_C}({\bf b})}
E^{{\ai},{\ai}^{*}}({\bf b},z)\,,
\end{equation} 
where $E^{{\ai},{\ai}^{*}}({\bf b},z)$ is the attenuation function
resulting from the elastic scattering in the strong fields of nucleons
before and after the conversion $\a \to \a^{*}$,
\begin{eqnarray}\label{N6}
& E^{{\ai},{\ai}^{*}}({\bf b},z) = &
\nonumber\\[0.5\baselineskip]
& \displaystyle = \exp\!\left\{ 
- A \! \int_{-\infty}^z \!\!\!\! \d z' \!\! \int \d^2 {\bf b'} 
                   \gamma^{\ai}     ( {\bf b}\!-\!{\bf b'} ) 
\rho({\bf b'},z')
- \!A \! \int_z^{\infty}  \!\!\!\! \d z' \!\! \int \d^2 {\bf b'} 
                   \gamma^{\ai^{*}} ( {\bf b}\!-\!{\bf b'} )
\rho({\bf b'},z')  \right\} \! . \;\; &
\end{eqnarray}
Taking into account the short-range nature of strong interactions,
(\ref{N6}) is reduced to
\begin{equation}\label{N7}
E^{{\ai},{\ai}^{*}}({\bf b},z) = \exp\!\left\{\!
-\frac{1}{2}\sigma_{\ai}^{\prime} A \, T_{-}({\bf b},z) 
-\frac{1}{2} {\sigma}_{\ai^{*}}^{\prime} A \, T_{+}({\bf b},z) 
\right\}\! ,
\end{equation}
where $T_{\pm}({\bf b},z)$ are the thickness functions,
\begin{equation}\label{N8}
T_{-}({\bf b},z) = \!\int_{-\infty}^z \!\!\!\! \d z' \rho({\bf b},z'), 
\quad
T_{+}({\bf b},z) = \!\int_z^{ \infty} \!\!\!\! \d z' \rho({\bf b},z').
\end{equation} 
Here we used the optical theorem $f(0) = \i k (4\pi)^{-1} \sigma'$,
\begin{equation}\label{N9}
\sigma' = \sigma_{tot} (1-\i\alpha) \,,
\qquad \alpha = \Re f(0) /\, \Im f(0) \,.
\end{equation}

The $\chi_{_C}({\bf b})$ in (\ref{N5}) is the Coulomb phase arising in
the case of charged incident particle. It is determined by the
same-type expression under the exponent in (\ref{N6}), but with
summation over protons and with the Coulomb profile functions. In
the case of a positively charged particle and a spherical nucleus, 
$\chi_{_C}({\bf b})$ is \cite{Glauber70}
\begin{equation}\label{N10}
\chi_{_C}({\bf b}) \;=\; 4 \pi Z \alpha \left[\ln(\mu b) 
\int_{0}^{b} T(b') \, b' \d b' + 
\int_{b}^{\infty} \! \ln(\mu b') \, T(b') \, b' \d b' \right] ,
\end{equation}
Here $Z$ is a number of protons, and
\begin{equation}\label{N11}
T({\bf b}) = \int_{-\infty}^{+\infty} \rho({\bf b},z) \; \d z \,.
\end{equation}
The $\mu$ in (\ref{N10}) is a dimensional parameter. It does not affect
the dependence of $\chi_{_C}$ on ${\bf b}$, and only defines the
additive constant part of the phase which is irrelevant. (We can put
$\mu =k$ for definiteness.)

In the case of conversion $\a \to \a^{*}$ due to Coulomb field,
summation in (\ref{N5}) goes over the charged protons. So in this case 
\begin{equation}\label{N12}
F_{c}^{\ai \to \ai^{*}}\!(\vec{\bf q}) = 
Z \int \! \d^3 \vec{\bf r} \; 
e^{\mbox{\scriptsize{i}} {\vec{\bf q} \vec{\bf r}}}
\left[ \int \d^3 \vec{\bf r}\,' \,
f_{c}^{\ai \to \ai^{*}} \! ( {\vec{\bf r}\!-\!\vec{\bf r}\,'} ) 
\rho(\vec{\bf r}\,') \right]
e^{\mbox{\scriptsize{i}} \chi_{_C}({\bf b})}
E^{{\ai},{\ai}^{*}}({\bf b},z)\,,
\end{equation} 
where $f_{c}^{\ai \to \ai^{*}}$ is the amplitude of inelastic Coulomb
scattering off a proton. 

In the case of conversion $\a \to \a^{*}$ due to strong interactions, it
is convenient to express the amplitude in terms of profile function for
elementary inelastic process. Omitting standard calculations, where the
short-range property of strong interactions is used, we arrive at 
\begin{equation}\label{N13}
F_{s}^{\ai \to \ai^{*}}\!(\vec{\bf q}) = 
A \, \frac{\i k}{2\pi} \int \d^3 \vec{\bf r} \; 
e^{\mbox{\scriptsize{i}} {\vec{\bf q} \vec{\bf r}}}
\left[ \int \d^2 {\bf b'} \,
\gamma_{s}^{\ai \to \ai^{*}}\!( {\bf b}\!-\!{\bf b'} ) \rho({\bf b'},z)
\right]e^{\mbox{\scriptsize{i}} \chi_{_C}({\bf b})}
E^{{\ai},{\ai}^{*}}({\bf b},z)\,.
\end{equation} 
Recall that $\vec{\,\bf q} = ({\bf q},q_{z})$, $\vec{\,\bf r} = ({\bf
b},z)$. The profile function $\gamma_{s}^{\ai \to \ai^{*}}\!( {\bf b} )$
is expressed through the inelastic elementary scattering amplitude
in the same way as in (\ref{N2}). 

\section{Form factors for direct pair-production}\label{sec3}

Based on (\ref{N12}) and (\ref{N13}), one can make a detailed definition
of amplitudes for various processes. In the initial work \cite{Faeldt43}
the inelastic process $\a \to \a^{*}$ was considered with the change
of spin of incident particle, of the type $K A \to K^{*} A$, in the
approximation of stable $\a^{*}$. It was also assumed that quantum
numbers exchanged are those of the photon. In this case the elementary
amplitude is proportional to the transverse component of the transfer
and may be written as
\begin{equation}\label{N14}
f_{c,s}^{\ai \to \ai^{*}}\!(\bar{\,\bf q})
= {\bf u q}\;\phi({\bar{\,\bf q}^2}) \,.
\end{equation} 
Here ${\bf u}$ is a vector in the impact-parameter plane,
$\phi({\bar{\,\bf q}^2})$ is proportional to $1/{\bar{\,\bf q}^2}$ in
the case of Coulomb forces and is finite at ${\bar{\,\bf q}^2} \to 0$ in
the case of strong interactions. Given this behavior, \cite{Faeldt43}
obtained the amplitudes for the coherent Coulomb and strong production.
In this section we determine the analogous formulas in the case of
direct production of a pair of particles. For definiteness we consider
the case of $K^{+} \pi^{0}$ production in the $K^{+}$ beam. However our
formulas will be valid for any process $\a A \to \a\c A$ with charged
$\a$ if the direct conversion $\a \to \a\c$ is possible due to the
chiral anomaly.

We start with the attenuation function. At first we note that at high
energies and small invariant mass of the $K\pi$ system the relative
angle between  scattering $K$ and $\pi$ is very small in the lab frame.
As a result, they do not have time to spread over long distances in the
impact-parameter plane during the passage of nucleus. In particular,
at 18 GeV incident $K$ and the invariant mass of $K\pi$ of the order of
the nominal mass of $K^{*}$, the $K$ and $\pi$ have time to spread at
the distance of order 1\% of the nucleus radius. This is much smaller
than the radius of nuclear forces. So the $K^{+} \pi^{0}$ system inside
the nucleus can be considered as a pair of unconnected particles moving
in parallel with a common impact parameter. The attenuation function of
such a system is formed from the products
\begin{equation}\label{N15}
\prod_{{j',j''=1}\atop{j'\not=j''\not=n}}^{A-1} \! 
\left[1\!-\! \theta(z_{j'}\!-\!z) 
\gamma^{\tiny\mbox{$K^{+}$}}_{j'}\!({\bf b}\!-\!{\bf b}_{j'}) \,
\right] 
\left[1\!-\!\theta(z_{j''}\!-\!z)\gamma^{\tiny\mbox{$\pi^{0}$}}_{j''}\!
({\bf b}\!-\!{\bf b}_{j''}) \, \right]
\end{equation}
instead of the products of
$\left[1\!-\!\theta(z_j\!-\!z)\gamma^{\ai^{*}}_j \! ({\bf b}\!-\!{\bf
b}_j ) \, \right]$ in formula (\ref{N1}). In the approximation
(\ref{N3}) and in the case of heavy nucleus, this leads to the
attenuation function
\begin{equation}\label{N16}
E^{\tiny\mbox{$K^{+}\!\!,K^{+}\!\pi^{0}$}}({\bf b},z) = 
\exp\!\left\{ -\frac{1}{2}\sigma_{\tiny\mbox{$K^{+}$}}^{\prime} 
A \, T_{-}({\bf b}) 
-\frac{1}{2} \left(\sigma^{\prime}_{\tiny\mbox{$K^{+}$}} \!+\!
{\sigma}^{\prime}_{\tiny\mbox{$\pi^{0}$}}\right) 
A \, T_{+}({\bf b},z) 
\right\}.
\end{equation}
Note that the cross sections in the second term in braces in (\ref{N16})
are determined with the momenta not equal to $k$, but defined by the
kinematics of the corresponding processes.

The averaged elementary amplitude of Coulomb scattering with direct
conversion $K^{+} \to K^{+}\pi^{0}$ due to the chiral anomaly is
\cite{Burtovoy13}  
\begin{equation}\label{N17}
f_{c}^{\tiny\mbox{$K^{+}\!\!\to\!K^{+}\!\pi^{0}$}}\!
(\vec{\,\bf q},\dots)
= \alpha \, \frac{{\bf c(\dots) q}}{\vec{\,\bf q}^2} \,.
\end{equation} 
Here $\bf c$ is a vector in the impact-parameter plane, and dots mean
kinematic variables that are additional to $\bar{\,\bf q}$. Calculating
Fourier with respect to $\bar{\,\bf q}$ and substituting the result into
(\ref{N12}), we get
\begin{eqnarray}\label{N18}
& F_{c}^{\tiny\mbox{$K^{+}\!\!\to\!K^{+}\!\pi^{0}$}}\!
(\vec{\,\bf q},\dots) = &
\nonumber\\[0.4\baselineskip]
& \displaystyle
= - \i Z\alpha \int \d^2 {\bf b} \, \d z \;
e^{\mbox{\scriptsize{i}} {\bf q b} + \mbox{\scriptsize{i}} q_{z} \! z} 
\left[ \frac{{\bf c(\dots) b}}{r^3} \int_0^r \d y \; y^2 \rho(y)
\right] e^{\mbox{\scriptsize{i}} \chi_{_C}({\bf b})}
E^{\tiny\mbox{$K^{+}\!\!,K^{+}\!\pi^{0}$}}({\bf b},z)\,, &
\end{eqnarray} 
where $r = \sqrt{{\bf b}^2+z^2}$. Calculating the angular integral, we
arrive at
\begin{equation}\label{N19}
F_{c}^{\tiny\mbox{$K^{+}\!\!\to\!K^{+}\!\pi^{0}$}}\!
(\vec{\bf q},\dots) = 
f^{\tiny\mbox{$K^{+}\!\!\to\!K^{+}\!\pi^{0}$}}_{c}\!
(\vec{\bf q},\dots) \;
\Phi_{c}^{\tiny\mbox{$K^{+}\!\!,K^{+}\!\pi^{0}$}}(\vec{\bf q})\,,
\end{equation} 
where $\Phi_{c}^{\tiny\mbox{$K^{+}\!\!,K^{+}\!\pi^{0}$}}$ is the Coulomb
form factor,\footnote{Note 
that similar formula (3.4) in \cite{Faeldt43} for the case of single
particle production contains inaccuracies: factor $\pi$ is lost and the
contribution of imaginary part of $\exp(\i q_{z} \! z)$ is ignored.}
\begin{eqnarray}\label{N20}
& \Phi_{c}^{\tiny\mbox{$K^{+}\!\!,K^{+}\!\pi^{0}$}}(\vec{\bf q})  = &
\nonumber\\[0.2\baselineskip]
& \displaystyle
= 2 \pi Z \frac{\vec{\,\bf q}^2}{q} 
\int_0^{\infty} \! b^2 \d b \, J_1 (b q) 
\int \!\d z \, e^{\mbox{\scriptsize{i}} q_{z}z} \,
\left[\frac{1}{r^3} \int_0^{r} \!\! \rho(y) \, y^2 \d y \right]
e^{\mbox{\scriptsize{i}} \chi_{_C}(b)} 
E^{\tiny\mbox{$K^{+}\!\!,K^{+}\!\pi^{0}$}}(b,z) , &
\end{eqnarray}
$J_1$ is the Bessel function. 

In the case of direct conversion $K^{+} \to K^{+} \pi^{0}$ due to strong
interactions, the elementary amplitude may include two contributions,
with normal and abnormal parity of the meson vertices,
\begin{equation}\label{N21}
f_{s}^{\tiny\mbox{$K^{+}\!\!\to\!K^{+}\!\pi^{0}$}}\!
(\vec{\,\bf q},\dots) = 
\phi_{s}^{n}({\vec{\,\bf q}^2},\dots) +
{\bf h(\dots) q} \, \phi_{s}^{a}({\vec{\,\bf q}^2},\dots) \,.
\end{equation}
Here dots mean additional kinematic variables like in (\ref{N17}), and
${\bf h}$ is a vector in the impact-parameter plane. In the case of
scattering at low energies, $\phi_{s}^{a}$ would imply exchanges by
isoscalar vector mesons ($\omega$, $\phi$) with anomalous vertex $KK \pi
V$. Similarly, $\phi_{s}^{n}$ would imply exchanges by isoscalar axial
mesons with normal-parity vertex $KK \pi A$. (Note that pseudoscalar
exchanges are forbidden as they imply spin flip of the nucleon, which
means the loss of the coherence.) At high energies the contributions in
the $t$-channel are reggeized. So $\phi_{s}^{n}$ and  $\phi_{s}^{a}$
have a form \cite{Regge,Irving}
\begin{equation}\label{N22}
\phi_{s}^{n,a}({\bar{\,\bf q}^2},\dots) = 
\frac{k}{4 \pi s} \sum_{i} \beta_{i}^{\,n,a}({\bar{\,\bf q}^2},\dots) \,
\frac{1 - e^{-i\pi\alpha_{i}}}{\sin \,\pi\alpha_{i}} \,
\left( s/s_0 \right)^{\alpha_{i}} ,
\end{equation}
where $\alpha_{i}$ are the Regge trajectories and $s$ is the
corresponding Mandelstam variable. The factor $k/(4\pi s)$ is due to the
normalization of the amplitude adopted in (\ref{N1}).

Substituting (\ref{N21}) into (\ref{N13}), with taking into account
(\ref{N2}), and considering that the range of strong interaction is much
smaller than the nuclear radius, we get
\begin{eqnarray}\label{N23}
& \displaystyle
F_{s}^{\tiny\mbox{$K^{+}\!\!\to\!K^{+}\!\pi^{0}$}}\!
(\vec{\,\bf q},\dots) = \phi_{s}^{n}(0,\dots) A \!
\int \! \d^2 {\bf b} \, \d z \;
e^{\mbox{\scriptsize{i}} {\bf q b} + \mbox{\scriptsize{i}} q_{z}z}
\rho(b,z) \,
e^{\mbox{\scriptsize{i}} \chi_{_C}(b)}
E^{\tiny\mbox{$K^{+}\!\!,K^{+}\!\pi^{0}$}}({\bf b},z)
& \nonumber\\[0.5\baselineskip]
&\displaystyle
+ \; \i \,\phi_{s}^{a}(0,\dots) A \! 
\int \! \d^2 {\bf b} \, \d z \;
e^{\mbox{\scriptsize{i}} {\bf q b} + \mbox{\scriptsize{i}} q_{z} z} \!
\left[ \frac{\bf h(\dots) b}{b} 
\frac{\partial \rho(b,z)}{\partial b} \right] \!
e^{\mbox{\scriptsize{i}} \chi_{_C}(b)}
E^{\tiny\mbox{$K^{+}\!\!,K^{+}\!\pi^{0}$}}({\bf b},z).
\end{eqnarray} 
After the calculation of angular integrals, we obtain
\begin{equation}\label{N24}
F_{s}^{\tiny\mbox{$K^{+}\!\!\to\!K^{+}\!\pi^{0}$}}\!
(\vec{\,\bf q},\dots) = 
\phi_{s}^{n}(0,\dots) \, 
\Phi_{s,n}^{\tiny\mbox{$K^{+}\!\!,K^{+}\!\pi^{0}$}}(\vec{\bf q}) +
{\bf h(\dots) q} \,
\phi_{s}^{a}(0,\dots) \, 
\Phi_{s,a}^{\tiny\mbox{$K^{+}\!\!,K^{+}\!\pi^{0}$}}(\vec{\bf q}),
\end{equation} 
where
\begin{equation}\label{N25}
\Phi_{s,n}^{\tiny\mbox{$K^{+}\!\!,K^{+}\!\pi^{0}$}} = 
2 \pi A \int_0^{\infty} b \d b \; J_0 (bq) 
\int \!\d z \, e^{\mbox{\mbox{\scriptsize{i}}} q_{z}z} 
\rho(b,z)  \,
e^{\mbox{\scriptsize{i}} \chi_{_C}(b)}
E^{\tiny\mbox{$K^{+}\!\!,\!K^{+} \pi^{0}$}}(b,z) \,,
\end{equation} 
\begin{equation}\label{N26}
\Phi_{s,a}^{\tiny\mbox{$K^{+}\!\!,K^{+}\!\pi^{0}$}} = 
- \frac{2 \pi A}{q}
\int_0^{\infty} b \d b \; J_1 (bq) 
\int \!\d z \, e^{\mbox{\mbox{\scriptsize{i}}} q_{z}z}
\frac{\partial \rho(b,z)}{\partial b}  \,
e^{\mbox{\scriptsize{i}} \chi_{_C}(b)}
E^{\tiny\mbox{$K^{+}\!\!,\!K^{+} \pi^{0}$}}(b,z) \,.
\end{equation}

\section{Unstable particles}\label{sec4}

Let us return to the case of conversion $\a \to \a^{*}$, and assume that
$\a^{*}$ is an unstable particle that can decay when passing the
nucleus. In this section we discuss how the description of section
\ref{sec2} must be changed in this case.  

First we note that as the decay is spontaneous, its probability can be
associated with the path length. So the probability that $\a^{*}$ being
produced at point $z$ reaches $z'$,~is  
\begin{equation}\label{N27}
w_{\ai^{*}}(z'-z) = \exp[-(z'-z)/l]\,,
\end{equation} 
where $l$ is the decay length. In the general case it may be considered
as a phenomenological parameter. In vacuum, $l = k/(m_{\ai^{*}}
\Gamma_{\ai^{*}})$. For simplicity we consider the case when $\a^{*}$
decays over a single channel ${\a}^{*} \to \b\c$. Then the probability
of occurrence $\b\c$ in the point $z'$ is 
\begin{equation}\label{N28}
w_{{\bi}{\ci}}(z'-z) = 1-w_{\ai^{*}}(z'-z)\,.
\end{equation} 

Next we note that a system that decays after production can be described
as a superposition of two states. In our case these are the quasi-stable
state $\a^{*}$ and the orthogonal state of the decay products $\b\c$.
Both states are taken with the weights $\sqrt{w_{\ai}}$ and
$\sqrt{w_{\bi\ci}} $, respectively. The amplitude of elastic scattering
of such a system is the sum of the amplitudes of elastic scattering of
each of its parts with the weights $w_{\ai}$ and $w_{\bi\ci}$.
Accordingly,
$\gamma^{\ai^{*}} ( {\bf b}\!-\!{\bf b'} )$ in (\ref{N6}) in this case
is replaced by 
\begin{equation}\label{N29}
w_{\ai^{*}}(z'\!-\!z) \, \gamma^{\ai^{*}}\!( {\bf b}\!-\!{\bf b'} ) +
w_{\bi\ci}(z'\!-\!z) \left[
\gamma^{\bi}_j ({\bf b}\!-\!{\bf b}_j ) + 
\gamma^{\ci}_j ({\bf b}\!-\!{\bf b}_j ) \right]  \,.
\end{equation} 
In the case of charged particles this construction leads to the same
Coulomb factor (\ref{N10}). However, the attenuation is different:
\begin{eqnarray}\label{N30}
& E^{{\ai},{\ai}^{*} \to \bi\ci}({\bf b},z) = &
\nonumber\\[0.5\baselineskip]
& \displaystyle
= \exp\!\left\{ 
-\frac{1}{2}\sigma_{\ai}^{\prime} A \, T_{-}({\bf b},z) 
-\frac{1}{2} {\sigma}_{\ai^{*}}^{\prime} 
A \, T_{+}^{\ai^{*}}({\bf b},z) 
-\frac{1}{2} ({\sigma}_{\bi}^{\prime}+{\sigma}_{\ci}^{\prime}) 
A \, T_{+}^{\bi\ci}({\bf b},z)
\right\} \! .&
\end{eqnarray} 
Here ${\sigma}_{\bi}^{\prime}$ and ${\sigma}_{\ci}^{\prime}$ are
determined at the momenta defined by the kinematics of the process, and
$T_{+}^{\mbox{x}}$ is a modified thickness function ($\mbox{\large x} =
\a^{*}$, $\b\c$),
\begin{equation}\label{N31}
T_{+}^{\mbox{x}}({\bf b},z) = 
\int_z^{ \infty}\!\!\! \d z'\,
w_{\mbox{x}}(z'\!-\!z)\rho({\bf b},z')\,.
\end{equation} 
It is readily seen that at $l \to \infty$ and $l \to 0$ formula
(\ref{N30}) gives the attenuation functions in the above cases of stable
${\a}^{*}$ and the direct conversion ${\a} \to \b\c$.

\begin{figure}[t]
\hbox{ \hspace*{140pt}
       \epsfxsize=0.35\textwidth \epsfbox{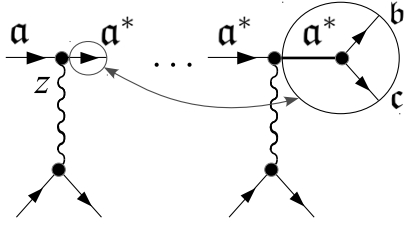}}
\caption{\small Formal replacement in the elementary amplitudes with the
production of particle $\a^{*}$ and its decay.
%\vspace*{-0.5\baselineskip} 
}
\label{Fig2}
\end{figure}

Now we define a place in the formula for the amplitude where the decay
vertex $\a^{*}\!\to \b\c$ has to make a contribution. The problem is
that since the decay is spontaneous the appropriate vertex may appear in
any place depending on where the decay occurred. However, on the other
hand, the decay vertex contributes necessarily together with the
propagator connecting it with the vertex of the last elastic $\a^{*}$
scattering and with the wave functions of the decay products of
$\a^{*}$. Since the averaged amplitude of elastic scattering is
proportional to the unit operator in spin variables and $\a^{*}$
momentum is constant in the leading approximation, the block of above
elements---the propagator of $\a^{*}$, the decay vertex and the wave
functions---may be formally attributed to the initial vertex, where
$\a^{*}$ was formed, i.e.~to the vertex of the conversion $\a \to
\a^{*}$. Simultaneously the wave function of $\a^{*}$ at the latter
vertex may be attributed to the vertex of the last elastic $\a^{*}$
scattering, see Fig.\ref{Fig2}. After performing these formal
manipulations, the result will be the replacement in formula (\ref{N5})
of the amplitude $f_{n}^{\ai \to \ai^{*}}$ by the amplitude $f_{n}^{\ai
\to \ai^{*} \to \bi\ci}$ for the cascade process $\a n \to \a^{*} n,
\a^{*} \to \b \c$, where $n$ is the nucleon of the nucleus. 

Thus, we arrive at the following formula for amplitude of the entire
coherent process:
\begin{eqnarray}\label{N32}
& F^{\ai \to \ai^{*} \to \bi\ci}(\vec{\bf q},\dots) = &
\nonumber\\[0.3\baselineskip]
&\displaystyle 
= \int \! \d^3 \vec{\bf r} \; 
e^{\mbox{\scriptsize{i}} {\vec{\bf q} \vec{\bf r}}} \!
\left[ \sum_{n=1}^{A} \! \int \! \d^3 \vec{\bf r}\,' \,
f_{n}^{\ai \to \ai^{*} \to \bi,\ci} 
( {\vec{\bf r}\!-\!\vec{\bf r}\,'},\dots) 
\rho(\vec{\bf r}\,') \! \right] \!
e^{\mbox{\scriptsize{i}} \chi_{_C}({\bf b})}
E^{{\ai},{\ai}^{*} \to \bi\ci}({\bf b},z)\,.&
\end{eqnarray} 
Similar replacements must be made in the formulas that follow
(\ref{N5}), including formulas for the form factors.

\section{Virtual particles}\label{sec5}

If system $\b\c$ in the cascade process $\a \to \a^{*}\!\to \b\c$ is
produced far from the $\a^{*}$ mass shell, then particle $\a^{*}$ must
be virtual, at least immediately before it is converted to the final
state. Unfortunately, a priori we do not known at what stage $\a^{*}$
becomes virtual, and optical analogues do not allow us to understand
where this occurs. For this reason, we turn to the field-theoretical
analysis of the coherent scattering carried out in monograph
\cite{AAV}. 

\begin{figure}[t]
\hbox{ \hspace*{125pt}
       \epsfxsize=0.4\textwidth \epsfbox{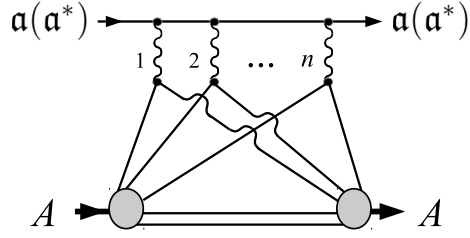}}
\caption{\small Elastic scattering of a fast particle off a
nonrelativistic ``soft'' system. The scattering off $n$ constituents of
the system is shown. The figure is taken from \cite{AAV}.
%\vspace*{-0.5\baselineskip}
}
\label{Fig3}
\end{figure}

The results of \cite{AAV} we need are summarized as follows. 
The scattering of a fast particle off a nonrelativistic ``soft'' system
(nucleus) consisting of $A$ constituents (nucleons) may be represented
as a convolution of the product of the wave functions of the ``soft''
system with the sum of the comb-shaped Green functions, see
illustration in Fig.\ref{Fig3}. The chord of the comb in the Green's
functions is made of the propagators of incident particle, and the teeth
are the propagators of intermediate particles that couple the incident
particle with the constituents of the ``soft'' system. Among~the
integration variables one can distinguish the virtualities of the chord
propagators. Further, it is assumed that through the chord a large
momentum flows, while through the teeth small transfers flow
(small-angle scattering that does not destroy the ``soft'' system).
Under these conditions, in integrals over the virtualities one can
distinguish a part in which the integration contour may be deformed in
such a way that only imaginary part of the propagator of the chord,
proportional to $\delta(k_j^2-m^2)$, makes contributions. Such
contributions break down the Green functions into the product of
elementary amplitudes. Moreover, their sum forms exactly Glauber
approximation. 

Simultaneously we know that Glauber approximation is the leading one.
From this we deduce that all contributions in the above consideration
with the off-shell chord propagators form a correction. At the level of
physical processes this means that if a virtual particle appears in the
chord, it immediately receives the necessary longitudinal momentum from
the constituents and becomes real. Otherwise it determines a correction
to the leading approximation.

On this basis we come to the following scenario. After multiple elastic
scattering, the incident particle $\a$ converts in the point $z$ into
the on-shell $\a^{*}$. Then a series of its elastic scattering follows.
In the last scattering, say in point $z_1$, $\a^{*}$ goes off the mass
shell, i.e.~becomes virtual $\tilde{\a}^{*}$, and then converts into
$\b\c$ before interacting with other nucleons. The latter system then
elastically scatter.

Since the last $\a^{*}$ scattering with the conversion $\a^{*} \!\to\!
\tilde{\a}^{*}$ is accompanied by the mass change, it should be
considered as an inelastic process. The generalization of formula
(\ref{N1}) to this case is as follows
\begin{eqnarray}\label{N33}
& \displaystyle 
F^{\ai \to \widetilde{\ai}^{*} \to \bi\ci} (\vec{\bf q},\dots) = -\!
\int  \d^2 {\bf b} \, \d z \;
e^{\mbox{\scriptsize{i}} {\bf q b} + \mbox{\scriptsize{i}} q_{z}\!z}
\left[ \sum_{n=1}^{A} \, \int \d^3 \vec{\bf r}\,' \,
f_{n}^{\ai \to \ai^{*}} 
( {\vec{\bf r}\!-\!\vec{\bf r}\,'} ) \,
\rho(\vec{\bf r}\,') \right] 
& \nonumber\\[-0.1\baselineskip]
& \displaystyle \times
\int \d z_1 \, \theta(z_1\!-\!z) \, 
e^{\mbox{\scriptsize{i}} q_{z_1}\!z_1} 
\left[ \sum_{n_1=1}^{A-1} \!
\int \! \d^3 \vec{\bf r}\,'_1 \, \frac{2\pi}{\i k} 
f_{n_1}^{\ai^{*} \to \widetilde{\ai}^{*} \to \bi\ci} 
( {{\bf b}\!-\!{\bf b}'_1,z_1\!-z'_1,\dots} ) \,
\rho(\vec{\bf r}\,'_1) \right] &
\nonumber\\[0.5\baselineskip]
& \displaystyle 
\times \;
e^{\mbox{\scriptsize{i}} \chi_{_C}({\bf b})} \,
\! E^{\ai,\ai^{*}\!,\bi\ci}({\bf b},z,z_1). & 
\end{eqnarray}
Here  $\bar{\bf q} = ({\bf q},q_{_{L}})$, and the common minus sign
arises due to the shadowing effect caused by the presence of two
inelastic processes. With large $k$ the longitudinal transfers $q_{z}$
and $q_{z_1}$ are 
\begin{equation}\label{N34}
q_{z}   = \frac{m_{\ai^{*}}^2 - m_{\ai}^2}{2k}\;, \qquad
q_{z_1} = \frac{m_{\widetilde{\ai}^{*}}^2  - m_{{\ai}^{*}}^2}{2k^{*}}\;,
\end{equation} 
where $m_{\widetilde{{\ai}}^{*}}^2=M_{\bi\ci}^2$, the invariant mass
squared of the $\b\c$. Thereby, the total longitudinal transfer is 
\begin{equation}\label{N35}
q_{_{L}} = \frac{m_{\widetilde{{\ai}}^{*}}^2 - m_{\ai}^2}{2k}\,.
\end{equation}
The attenuation function in (\ref{N33}) is
\begin{eqnarray}\label{N36}
& E^{\ai,\ai^{*}\!,\bi\ci}({\bf b},z,z_1)=&
\nonumber\\[0.6\baselineskip]
& \displaystyle = \exp\!\left\{ 
-\frac{1}{2}\sigma_{\ai}^{\prime} A \, T_{-}({\bf b},z) 
-\frac{1}{2} {\sigma}_{\ai^{*}}^{\prime} 
A \, T({\bf b}, z, z_1) 
-\frac{1}{2} ({\sigma}_{\bi}^{\prime}+{\sigma}_{\ci}^{\prime}) 
A \, T_{+}({\bf b},z_1) \right\}  ,&
\end{eqnarray} 
with $T_{-,+}({\bf b},z)$ defined in (\ref{N8}), and $T({\bf b}, z,
z_1)$ is
\begin{equation}\label{N37}
T({\bf b},z,z_1) = 
\int_z^{z_1}\!\!\! \d z'\,\rho({\bf b},z')\,.
\end{equation} 
Recall that ${\sigma}_{\bi}^{\prime}$ and ${\sigma}_{\ci}^{\prime}$ are
determined at the momenta defined by the kinematics of the process.

\begin{figure}[t]
\hbox{ \hspace*{140pt}
       \epsfxsize=0.35\textwidth \epsfbox{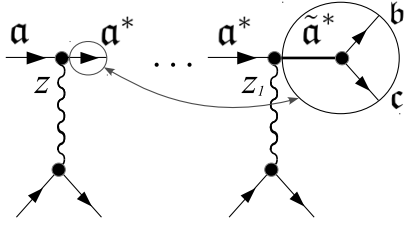}}
\caption{\small The same as in Fig.\ref{Fig2} with virtual
$\tilde{\a}^{*}$.
%\vspace*{-0.5\baselineskip} 
}
\label{Fig4}
\end{figure}

The $f_{n_1}^{\ai^{*} \to \widetilde{\ai}^{*} \to \bi\ci}$ in
(\ref{N33}) is the elementary amplitude of the cascade process. It
includes, in particular, the propagator of virtual $\widetilde{\a}^{*}$
and the decay vertex with wave functions. Repeating the reasoning of
section \ref{sec4}, we can formally attribute the above elements to the
vertex $\a \to \a^{*}$, and simultaneously attribute the wave function
of $\a^{*}$ at the latter vertex to the vertex $\a^{*} \to
\widetilde{\a}^{*}$, see illustration in Fig.\ref{Fig4}. In doing so, we
leave the exponents with the phase shifts in the former places, and we
do not change the attenuation function. As a result we arrive at the
equivalent formula,
\begin{eqnarray}\label{N38}
& \displaystyle 
F^{\ai \to \widetilde{\ai}^{*} \to \bi\ci} (\vec{\bf q},\dots) = -\!
\int  \d^2 {\bf b} \, \d z \;
e^{\mbox{\scriptsize{i}} {\bf q b} + \mbox{\scriptsize{i}} q_{z}\!z}
\left[ \sum_{n=1}^{A} \, \int \d^3 \vec{\bf r}\,' \,
f_{n}^{\ai \to \widetilde{\ai}^{*} \to \bi\ci} 
( {\vec{\bf r}\!-\!\vec{\bf r}\,',\dots} ) \,
\rho(\vec{\bf r}\,') \right] 
& \nonumber\\
& \displaystyle \times
\int \d z_1 \, \theta(z_1\!-\!z) \, 
e^{\mbox{\scriptsize{i}} q_{z_1}\!z_1} 
\left[ \sum_{n_1=1}^{A-1} \!
\int \! \d^3 \vec{\bf r}\,'_1 \, 
\frac{2\pi}{\i k} f_{n_1}^{\ai^{*} \to \ai^{*}} \!
( {{\bf b}\!-\!{\bf b}'_1,z_1\!-z'_1} ) \,
\rho(\vec{\bf r}\,'_1) \right] &
\nonumber\\[0.5\baselineskip]
& \displaystyle 
\times \;
e^{\mbox{\scriptsize{i}} \chi_{_C}({\bf b})} \,
\! E^{\ai,\ai^{*}\!,\bi\ci}({\bf b},z,z_1)\,, & 
\end{eqnarray}
where $f_{n}^{\ai \to \widetilde{\ai}^{*} \to \bi\ci}$ and
$f_{n_1}^{\ai^{*} \to \ai^{*}}$ are the elementary amplitudes of the
cascade process and the elastic scattering. Their Fourier by $z$ and
$z_1$ lead to formally improper longitudinal transfers since we did not
rearrange the phase factors. However, this does not affect the full
amplitude since the elementary inelastic vertices do not depend on the
longitudinal transfers. The dependence on them is contained in the
$t$-channel propagators that were not involved in our above
manipulations.

Based on (\ref{N38}), we can further refine the definition of the
amplitude. Namely, we note that if $\a^{*}$ is a charged particle, then
$f_{n_1}^{\ai^{*} \to \ai^{*}}$ includes two contributions, due to
strong and Coulomb interactions. In the former case, in view of the
short-range nature of strong interactions, the integral $\d^3 \bar{\bf
r}\,'_1$ in (\ref{N38}) is reduced to
\begin{equation}\label{N39}
V_s ({\bf b},z_1) = 
\frac{1}{2}\sigma_{\ai^{*}}^{\prime} \, \rho({\bf b},z_1)\,.
\end{equation} 
In the case of a long-range Coulomb interaction the integral $\d^3
\bar{\bf r}\,'_1$ is reduced to the one-fold integral
\begin{equation}\label{N40}
V_c ({\bf b},z_1) =
4\pi\i\alpha \left[ \frac{1}{r_1} \int_0^{r_1} \rho(y) y^2 \d y + 
\int_{r_1}^{\infty} \!\rho(y) y \d y \right] .
\end{equation} 
Here $r_1 = \sqrt{{\bf b}^2+z_1^2}$, and ``$\i$'' appears because we
take the Coulomb amplitude in the Born approximation, which is real. So,
instead of (\ref{N38}) we get
\begin{eqnarray}\label{N41}
& \displaystyle 
F^{\ai \to \widetilde{\ai}^{*} \to \bi\ci} (\vec{\bf q},\dots) = -\!
\int \d^2 {\bf b} \, \d z \;
e^{\mbox{\scriptsize{i}} {\bf q b} + \mbox{\scriptsize{i}} q_{z}\!z}
\left[ \sum_{n=1}^{A} \; \int \d^3 \vec{\bf r}\,' \,
f_{n}^{\ai \to \widetilde{\ai}^{*} \to \bi\ci} 
( {\vec{\bf r}\!-\!\vec{\bf r}\,',\dots} ) \,
\rho(\vec{\bf r}\,') \right]
& \nonumber\\[0.5\baselineskip]
& \displaystyle 
\times \int_{z}^{\infty} \! \d z_1 \, 
e^{\mbox{\scriptsize{i}} q_{z_1}\!z_1} 
\Bigl[ (A\!-\!1) \, V_s ({\bf b},z_1) + (Z\!-\!1) \, V_c ({\bf b},z_1)
\Bigr] \;
e^{\mbox{\scriptsize{i}} \chi_{_C}({\bf b})} \,
\! E^{\ai,\ai^{*}\!,\bi\ci}({\bf b},z,z_1)\,. &
\end{eqnarray} 
Outside the nucleus $V_c$ absolutely dominates. Inside, $V_c$ and $V_s$
may be comparable despite the factor $\alpha$ in (\ref{N40}) because of
the factor $4\pi$ and the big size of the nucleus. Really, the absolute
value of $\sigma^{\prime}$ is usually about several tens mb, i.e.
several fm$^2$. At distances close to the radius of the nucleus, the
square brackets in (\ref{N40}) is roughly estimated as $\rho R^2/3$ with
$R \approx A^{1/3}$fm. Hence $|V_c/V_s| \approx (8/3) \pi \alpha
A^{2/3}$ which is about 1 if $A=63$, the case of Cu.

Calculating the angular integral in (\ref{N41}), we arrive at the final
results. So, in the case of Coulomb scattering, repeating the
calculations of sect.~\ref{sec3} we get
\begin{equation}\label{N42}
F_{c}^{\ai \to \widetilde{\ai}^{*} \to \bi\ci} (\bar{\bf q},\dots\!) = 
f^{\ai \to \widetilde{\ai}^{*} \to \bi\ci}_{c } (\bar{\bf q},\dots\!) \,
\Phi_{c}^{\ai,\ai^{*}\!,\bi\ci} (\bar{\bf q}),
\end{equation} 
where $\Phi_{c}^{\ai,\ai^{*}\!,\bi\ci}$ is the Coulomb form factor,
\begin{eqnarray}\label{N43}
&  \displaystyle
\Phi_{c}^{\ai,\ai^{*}\!,\bi\ci} (\vec{\bf q}) = -
2\pi Z \frac{\vec{\,\bf q}^2}{q} \int_0^{\infty}\! b^2 \d b \, J_1 (bq)
\int \!\d z \, e^{\mbox{\scriptsize{i}} q_{z}z} \,
\frac{1}{r^3} \int_0^{r} \!\! \rho(y) \, y^2 \d y &
\nonumber\\[0.5\baselineskip]
&  \displaystyle
\times \int_{z}^{\infty} \!\! \d z_1  \, 
e^{\mbox{\scriptsize{i}} q_{z_1} z_1} \, 
\Bigl[ (A\!-\!1) \, V_s + (Z\!-\!1) \, V_c \Bigr] \,
e^{\mbox{\scriptsize{i}} \chi_{_C}(b)}
E^{\ai,\ai^{*}\!,\bi\ci}(b,z,z_1)\,.
& 
\end{eqnarray}
If the conversion $\a \to \a^{*}$ occurs in the strong field of the
nucleus, then the elementary and the full amplitudes are given by
(\ref{N21}) and (\ref{N24}), respectively, with the obvious changing in
superscripts, and with the form factors
\begin{eqnarray}\label{N44}
&  \displaystyle
\Phi_{s,n}^{\ai,\ai^{*}\!,\bi\ci} = -
2\pi A \, \int_0^{\infty} b \d b \; J_0 (bq) \,
\int \!\d z \, e^{\mbox{\mbox{\scriptsize{i}}} q_{z}z} \,
\rho(b,z) &
\nonumber\\[0.2\baselineskip]
&  \displaystyle
\times \int_{z}^{\infty} \!\! \d z_1  \, 
e^{\mbox{\scriptsize{i}} q_{z_1} z_1} \, 
\Bigl[ (A\!-\!1) \, V_s + (Z\!-\!1) \, V_c \Bigr] \,
e^{\mbox{\scriptsize{i}} \chi_{_C}(b)}
E^{\ai,\ai^{*}\!,\bi\ci} (b,z,z_1)\,,
& 
\end{eqnarray} 
\begin{eqnarray}\label{N45}
&  \displaystyle
\Phi_{s,a}^{\ai,\ai^{*}\!,\bi\ci} = 2\pi A \, \frac{1}{q}
\int_0^{\infty} b \d b \; J_1 (bq) \,
\int \!\d z \, e^{\mbox{\mbox{\scriptsize{i}}} q_{z}z} 
\frac{\partial \rho(b,z)}{\partial b} &
\nonumber\\[0.2\baselineskip]
&  \displaystyle
\times \int_{z}^{\infty} \!\! \d z_1  \, 
e^{\mbox{\scriptsize{i}} q_{z_1} z_1} \, 
\Bigl[ (A\!-\!1) \, V_s + (Z\!-\!1) \, V_c \Bigr] \,
e^{\mbox{\scriptsize{i}} \chi_{_C}(b)}
E^{\ai,\ai^{*}\!,\bi\ci}(b,z,z_1)\,.
& 
\end{eqnarray} 
In the particular case of scattering $K^{+} A \to K^{+} \pi^{0} A$ via
virtual ${K}^{*+}$ in the $s$-channel, one should substitute
$K^{+}\!$, $K^{*+}\!$, $K^{+}\!\pi^{0}$ for $\a,\a^{*}\!,\b\c$ in
(\ref{N42})--(\ref{N45}).

In the end of this section, we consider another scheme of the process
with virtual particles, namely $\a \to \b \widetilde{\u} \to \b\c$, see
examples in Fig.\ref{Fig1}c,d. In this scheme, after the $\a$ multiple
scattering a spontaneous transition $\a \to \b \tilde{\u}$ occurs with
the production of real $\b$ and virtual $\tilde{\u}$. Then $\tilde{\u}$
scatters off a nucleon in the point $z_1$, and becomes real $\u$. The
$\b$ begins elastic scattering at $z_1$, as well. After a series of
elastic scatterings, $\u$ converts to $\c$ in the point $z$, and $\c$
elastically scatters. 

The generalization of the above formulas to this scenario is obvious:
\begin{eqnarray}\label{N46}
& \displaystyle 
F^{\ai \to \bi\widetilde{\ui} \to \bi\ci} (\vec{\bf
q},\dots) = -\!\!
\int \d^2 {\bf b} \, \d z_1 \,
e^{\mbox{\scriptsize{i}}{\bf q b}+\mbox{\scriptsize{i}} q_{z_1}\!z_1} \!
\int_{z_1}^{\infty} \!\! \d z \, e^{\mbox{\scriptsize{i}} q_{z}\!z}
\Bigl[ (A\!-\!1) \, V_s + (Z\!-\!1) \, V_c \Bigr]  &
\nonumber\\[0.3\baselineskip]
& \displaystyle \!\!\!\!\!
\times 
\left[ \sum_{n=1}^{A} \!\int \d^3 \vec{\bf r}\,' \!
f_{n}^{\ai \to \bi\widetilde{\ui} \to \bi\ci} 
( {\vec{\bf r}\!-\!\vec{\bf r}\,',\dots} ) 
\rho(\vec{\bf r}\,') \right] 
e^{\mbox{\scriptsize{i}} \chi{_C}({\bf b})} \,
\! E^{\ai,\bi,\ui,\ci}({\bf b},z_1,z)\,. &
\end{eqnarray} 
Notice, here $z > z_1$ and $V_s$ includes $\sigma_{\ui}^{\prime}$
instead of $\sigma_{\ai^{*}}^{\prime}$ in (\ref{N41}). The longitudinal
momenta in (\ref{N46}) are as follows
\begin{equation}\label{N47}
q_{z_1} = \frac{m_{{\ui}}^2
- m_{\widetilde{\ui}}^2}{2k_{\widetilde{{\ui}}}}\;, \qquad
q_{z}   = \frac{m_{\ci}^2 -
m_{\ui}^2}{2k_{\ui}}\;,
\end{equation} 
where $k_{\widetilde{{\ui}}}\,$, $m^2_{\widetilde{\ui}}$ are determined
by the kinematics of the process. The attenuation function is
\begin{eqnarray}\label{N48}
& \displaystyle 
E^{\ai,\bi,\ui,\ci}({\bf b},z_1,z) =
\exp\!\left\{ 
-\;\frac{1}{2}\sigma_{\ai}^{\prime} A \, T_{-}({\bf b},z_1) \right.&
\nonumber\\[0.5\baselineskip]
& \displaystyle \left.
-\;\frac{1}{2}\sigma_{\bi}^{\prime} A \, T_{+}({\bf b},z_1)
-  \frac{1}{2} {\sigma}_{\ui}^{\prime} 
A \, T({\bf b}, z_1, z) 
-  \frac{1}{2}\sigma_{\ci}^{\prime} A \, T_{+}({\bf b},z)
\right\} .
\end{eqnarray} 

Based on (\ref{N46}) and acting by analogy, the formulas for the
processes Fig.\ref{Fig1}c,d can be easily written. Because of the
bulkiness, we do not give them here.

\section{Discussion and conclusion}\label{sec6}

In this paper we proceeded from the provision about instantaneous
particle formation. Actually this is a common place in Glauber theory.
Nevertheless, this is a model assumption, which in a strict sense is not
quite correct as the complete formation of particles takes a time. In
particular, when converting $\a \to \a^{*}$ this time is of order
$q_z^{-1}$ in the lab frame. During this time a pre-particle $\a^{*}$,
before it becomes a full-fledged particle, passes a distance much larger
than the nucleus size. For example, $q_z^{-1} \approx 13$ fm in the case
of $K^{*}$ formation in the 18 GeV $K$ beam, while e.g.~the Cu radius is
about 4.4~fm. In the general case, if $q_z^{-1}$ is close to or smaller
than the nucleus size, then $q_z$ is too large, incompatible with the
nucleus integrity. Hence, the $\sigma'$ in the attenuation functions
actually corresponds to the pre-particle formations rather than real
particles. However, in the case of resonances $\sigma'$ is practically
determined within the Glauber theory framework. This actually eliminates
the problem. In other cases, in view of large $q_z^{-1}$ the mentioned
effect means mainly parametric change of $\sigma'$. This can always be
taken into account. Thus, the assumption of instantaneous particle
formation should not cause severe problems.

Turning to the results of this paper, we recall that~our goal was to
determine the form factors for the coherent scattering of fast particles
off heavy nuclei with the production of pairs of particle in the final
state. In the case of the direct pair production we were based on
combining the results for coherent inelastic one-particle scattering and
elastic scattering of composite systems. If the pair is produced via
intermediate particles, the solution depends on whether they are real
unstable or virtual ones. In the former case the form factors include
the modified thickness functions weighted with the probabilities for
survival and decay of the unstable particle. In the case of virtual
intermediate particles, we found that they can be virtual only before
their first interaction with nucleons of the nucleus or after the last
interaction just before they decay to real particles. At other stages
fast particles can exist inside the nucleus only as real ones, at least
in the leading approximation. The description of the conversion from
real to virtual state and vice versa requires an introduction of
additional inelastic-scattering vertex, through which the longitudinal
component of the momentum is transferred when the virtuality of the
particle changes.

Concerning the specific reaction $K^{+}\! A \to K^{+}\!\pi^{0} A$ which
is currently being studied at 18 GeV $K^{+}$ \cite{OKA}, we obtained the
required formulas, but we did not make quantitative estimates. We notice
only that in this case our result about unstable particles is of no
practical importance, since the decay length of the $K^{*}$ is too
long. Namely, it is approximately 85 fm which is to be compared with the
radius of the target nucleus of 4.4 fm. So the $K^{*}$ should be
considered as a stable particle. (At the same time, our results about
contributions of virtual particles are fully relevant.) However, at
lower energies and wider resonances the situation may change. For
instance, in the coherent scattering $\pi A \to \pi \pi A$ at 2 GeV the
decay length of intermediate $\rho$ is about 3.4~fm. This means that
effect of instability of $\rho$ must be taken into consideration for
most heavy nuclei. The same is the case in other reactions at similar
energies with $\rho$ production inside the nucleus. 

In general, our study supplements the existing description of the
coherent inelastic scattering off heavy nuclei, traced to
\cite{Faeldt43}. The obtained results can be directly applied in the
experimental study of the coherent scattering $K A \to K \pi A$ and $\pi
A \to \pi\pi A$. With appropriate modifications, our results may be
applied to other reactions with the pair production off nuclei via
intermediate contributions.

\begin{flushleft}
\bf Acknowledgement
\end{flushleft}
The author is grateful to V.F.Obraztsov for proposing the problem of
determining the form factors for coherent pair-production of particles
off heavy nucleus, and to V.S.Burtovoy for useful discussions.

\end{document}